\documentclass[iop]{emulateapj}
\usepackage{amsmath}

\begin{document}

\title{Modeling the Infrared Emission in Cygnus A}

\author{G. C. Privon\altaffilmark{1,}\altaffilmark{2}}

\author{S. A. Baum\altaffilmark{1,}\altaffilmark{3}}

\author{C. P. O'Dea\altaffilmark{4,}\altaffilmark{5}}

\author{J. Gallimore\altaffilmark{6,}\altaffilmark{7}}

\author{J Noel-Storr\altaffilmark{1}}

\author{D. J. Axon\altaffilmark{3,}\altaffilmark{8}}

\author{A. Robinson\altaffilmark{3}}

\altaffiltext{1}{Chester F. Carlson Center for Imaging Science, Rochester Institute of Technology, Rochester, NY 14623}
\altaffiltext{2}{Department of Astronomy, University of Virginia, Charlottesville, VA 22904}
\altaffiltext{3}{Radcliffe Institute for Advanced Study. Cambridge, MA 02138}
\altaffiltext{4}{Department of Physics, Rochester Institute of Technology, Rochester, NY 14623}
\altaffiltext{5}{Harvard-Smithsonian Center for Astrophysics. Cambridge, MA 02138}
\altaffiltext{6}{Department of Physics, Bucknell University, Lewisburg, PA 17837}
\altaffiltext{7}{National Radio Astronomy Observatory, Charlottesville, VA, 22904}
\altaffiltext{8}{School of Mathematical and Physical Sciences, University of Sussex, Falmer, Brighton BN1 9RH, UK 4}

\begin{abstract}
We present new Spitzer IRS spectroscopy of Cygnus A, one of the most luminous radio sources in the local universe. Data on the inner $20\arcsec$ are combined with new reductions of MIPS and IRAC photometry as well as data from the literature to form a radio through mid-infrared spectral energy distribution (SED). This SED is then modeled as a combination of torus reprocessed active galactic nucleus (AGN) radiation, dust enshrouded starburst, and a synchrotron jet. This combination of physically motivated components successfully reproduces the observed emission over almost 5 dex in frequency. The bolometric AGN luminosity is found to be $10^{12}$ L$_{\odot}$ ($90$\% of $L_{IR}$), with a clumpy AGN-heated dust medium extending to $\sim130$ pc from the supermassive black hole. Evidence is seen for a break or cutoff in the core synchrotron emission. The associated population of relativistic electrons could in principle be responsible for some of the observed X-ray emission though the synchrotron self-Compton mechanism. The SED requires a cool dust component, consistent with dust reprocessed radiation from ongoing star formation. Star formation contributes at least $6\times10^{10}$ $L_{\odot}$ to the bolometric output of Cygnus A, corresponding to a star formation rate of $\sim10$ $M_{\odot}$ yr$^{-1}$.

\end{abstract}


\section{Introduction}
As one of the nearest powerful radio-loud active galactic nuclei (AGN), Cygnus A provides an excellent laboratory to study the environment and activity of powerful AGN. The luminosity of the AGN in Cygnus A (which is predominantly expressed in the infrared) is high enough to classify it as quasar \citep[e.g.,][]{Djorgovski91}. While not seen in total intensity, broad H$\alpha$ is seen in polarized light \citep{Ogle97}, lending support for the existence of an obscured broad-line region (BLR). The polarized broad lines were detected within the ionization cone seen by \citet{Jackson96}, giving them a possible scattering origin in the NLR. 

The host galaxy of Cygnus A is a cD elliptical near the center of a cluster which appears to be undergoing a merger with cluster of similar size \citep{Ledlow05}. \citet{Tadhunter03} obtained Hubble Space Telescope (HST) Space Telescope Imaging Spectrograph (STIS) spectra of the nuclear region. Stepping the slit across the nucleus, a velocity gradient indicative of rotation around the radio axis was observed. Modeling the velocity as due to the potential of a supermassive black hole (SMBH) and stellar mass distribution (measured from a $1.6\mu$m NICMOS image) gives a SMBH mass of $2.5\pm0.7\times10^{9}$ $M_{\odot}$. This measurement is consistent with black-hole-mass--host-galaxy relations \citep[e.g.,][]{Magorrian98,Ferrarese00,Gebhardt00,Gultekin09}.

The large scale radio morphology shows prominent hotspots, lobes, and a radio core. Both a jet and counterjet are visible in Very Large Array (VLA) and Very Long Baseline Array (VLBA) observations \citep[e.g.,][]{Sorathia96}. Due to the edge-brightened morphology it is classified as an FR II source \citep[Fanaroff-Riley Type II;][]{Fanaroff74}.

X-ray observations of Cygnus A are consistent with the presence of a hidden quasar. Chandra ACIS observations from $0.7$ to $9$ keV by \citet{Young02} show the hard X-ray flux to be peaked at a location consistent with that of the radio core and unresolved (less than $0.\arcsec 4$ in size, determined by comparison with a model PSF). Higher energy INTEGRAL observations show emission between $20$ and $100$ keV \citep{Beckmann06}, although there may be some contamination from intracluster gas.  This hard X-ray emission is likely due to accretion disk emission Comptonized in the AGN corona, although the UV/optical emission is obscured.

In addition to the AGN activity, HST imaging has also revealed star formation in the central region of Cygnus A, which began $< 1$ Gyr ago \citep{Jackson98}. It is located in a $4$ kpc ring around the nucleus, oriented orthogonal to the radio axis.

Based on adaptive optics observations showing a secondary point source near the nucleus, \citet{Canalizo03} suggest that Cygnus A may be in the late stages of a merger event. This merger event and related accretion may be related to the current epoch of nuclear activity. 

A near-infrared spectrum presented by \citet{Bellamy04} showed complicated emission line properties suggesting an infalling molecular cloud, consistent with the \citet{Canalizo03} picture of a minor merger.  The H$_2$ lines were seen in several components, both redshifted and blueshifted relative to the systemic velocity, interpreted as emission from a rotating torus. The observed near-infrared line ratios are consistent with excitation by X-rays (from the AGN), while likely ruling out shocks as a possible excitation method.

\begin{figure}
\includegraphics[angle=270,width=0.5\textwidth]{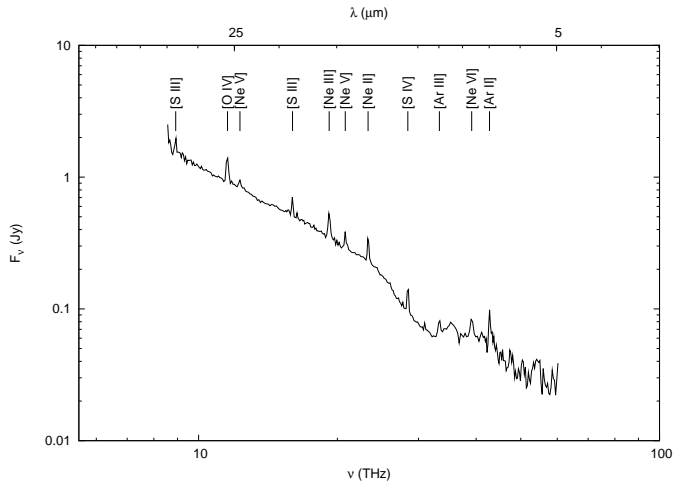}
\caption{Spitzer IRS spectrum for Cygnus A. Extracted in a $20\arcsec$ circular aperture from spectral mapping data.}
\label{fig:irs}
\end{figure}

For a more detailed summary of Cygnus A, including properties of the larger host galaxy and environment, see the review by \citet{Carilli96}.

While the bulk of the infrared emission is likely related to the presence of an AGN, the star formation can contribute to the infrared emission as well. Additionally, the bolometric luminosity of the AGN is uncertain. Estimates have been made using template spectral energy distributions (SEDs) and X-ray observations. However, a significant portion of the bolometric luminosity comes out in the infrared, via dust reprocessing of the UV/optical continuum. An accurate determination of the bolometric luminosity then requires an understanding of the contributions to the observed infrared luminosity. 

In this paper, we present modeling of the SED of Cygnus A. To accomplish this we combined new infrared measurements obtained by the Spitzer Space Telescope's Infrared Spectrograph \citep[IRS;][]{Werner04} with existing Spitzer observations with the Imaging Array Camera \citep[IRAC;][]{Fazio04} and the Multiband Imaging Photometer for SIRTF \citep[MIPS;][]{Rieke04}, and measurements of the radio core from the literature to construct a radio through infrared ($\sim2-10^{5}$ GHz; $4-10^{5}$ $\mu$m) SED of the inner regions of Cygnus A. The resulting SED was subsequently modeled using components intended to replicate the physical processes likely to produce the observed emission. Using the results of the fitting we have been able to decompose the infrared emission and determine the bolometric luminosity of the AGN in Cygnus A. This in turn also provides an estimate of the star formation rate.

The paper is organized as follows: in Section \ref{sec:data}, we discuss the new Spitzer IRS observations, new reductions of Spitzer IRAC and MIPS data, as well as the data from the literature. In Section \ref{sec:modeling}, we describe the components used to model the SED, for which the results are described in Section \ref{sec:results}. We conclude with a general discussion in Section \ref{sec:discussion} and a summary in Section \ref{sec:summary}.

\section{Data}
\label{sec:data}

New observations were combined with data compiled from multiple archives and new reductions in order to obtain a SED covering almost five orders of magnitude in frequency. Our analysis is focused on the properties of the continuum emission in this system. Below we discuss the reduction and analysis of the new Spitzer IRS observations and the re-reduction of archival IRAC and MIPS observations

\subsection{Spitzer IRS Observations}

Cygnus A was observed (PI: Baum) in ``mapping mode'' using the low resolution mode of IRS on board the Spitzer Space Telescope. Both the short- and long-wavelength slits were stepped across the source in increments of half the slit width. After pipeline calibration at the Spitzer Science Center, the observations were combined into a spectral data cube using the \emph{Cube Builder for IRS Spectra Maps} \citep[CUBISM;][]{Smith07b}. From this data cube, a spectrum was extracted in a $20\arcsec$ aperture (Figure \ref{fig:irs}).

\subsubsection{Removal of Mid-infrared Emission Lines}

The IRS spectrum was fit using PAHFIT \citep{Smith07a} to measure and remove contributions from narrow emission lines. Integrated line fluxes and widths are provided in Table \ref{table:mirlines}. While a study of the emission line properties is beyond the scope of this paper, we note the detection of multiple high ionization lines such as [O IV], [Ne V] and [Ne VI]. These are all consistent with the presence of an AGN \citep[][and references therein]{Genzel98,Armus07}. Of particular note are [Ne V] and [Ne VI] which require the presence of ionizing photons of at least 97.1 and 125.8 eV respectively. All the measured emission lines in Table \ref{table:mirlines} were subtracted from the IRS spectrum before fitting the SED.

\begin{deluxetable}{lcc}
\tablecaption{Mid-infrared Emission Line Properties}
\tablehead{\colhead{Line} & \colhead{Flux} & \colhead{FWHM}\\
\colhead{} & \colhead{\emph{($\times 10^{-13}$ erg s$^{-1}$ cm$^{-2}$)}} & \colhead{\emph{($\mu$m)}}}
\startdata
[Ar II] 6.9	& $1.93\pm0.08$	& 0.11 \\
{}[Ne VI] 7.6	& $1.37\pm0.09$	& 0.11 \\
{}[Ar III] 8.9	& $1.11\pm0.04$ & 0.11 \\
{}[S IV	10.5	& $1.55\pm0.03$ & 0.09 \\
{}[Ne II] 12.8	& $2.67\pm0.03$ & 0.11 \\
{}[Ne V] 14.3	& $1.51\pm0.04$ & 0.10 \\
{}[Ne III] 15.6	& $4.13\pm0.04$ & 0.15 \\
{}[S III] 18.7	& $2.42\pm0.06$ & 0.13 \\
{}[Ne V] 24	& $2.51\pm0.06$ & 0.37 \\
{}[O IV] 25.9	& $4.58\pm0.08$ & 0.32 \\
{}[S III] 33	& $3.33\pm0.08$ & 0.31 \\
\hline
H$_2$ S(7)	& $0.944\pm0.130$ & 0.06 \\
H$_2$ S(5)	& $0.853\pm0.112$ & 0.06 \\
H$_2$ S(3)	& $0.286\pm0.030$ & 0.09 \\
H$_2$ S(2)	& $0.333\pm0.025$ & 0.11 \\
H$_2$ S(0)	& $0.128\pm0.038$ & 0.31
\enddata
\tablecomments{Only detections $>3\sigma$ are listed. Errors quoted from PAHFIT output.}
\label{table:mirlines}
\end{deluxetable}

\subsubsection{Dust Features}

In addition to fitting the emission lines, we have also fit the mid-infrared dust features using PAHFIT. Table \ref{table:dustfeatures} shows the integrated flux and profile FWHM based on the PAHFIT output. In contrast with the emission lines, the dust features were not removed as they are expressed in the starburst models.

\begin{deluxetable}{lcc}
\tablecaption{Mid-infrared Dust Features}
\tablehead{\colhead{$\lambda$} & \colhead{Flux} & \colhead{FWHM}\\
\colhead{($\mu$m)} & \colhead{\emph{($\times 10^{-13}$ erg s$^{-1}$ cm$^{-2}$)}} & \colhead{\emph{($\mu$m)}}}
\startdata
$6.2$   &       $1.76\pm0.23$ &       $0.19$ \\
$6.7$   &       $5.99\pm0.49$ &       $0.47$ \\
$7.4$   &       $22.3\pm0.8$ &       $0.94$ \\
$7.8$   &       $2.26\pm0.32$ &       $0.42$ \\
$8.3$   &       $5.50\pm0.23$ &       $0.42$ \\
$8.6$   &       $4.42\pm0.20$ &       $0.34$ \\
$11.3$  &       $0.685\pm0.101$ &       $0.36$ \\
$12.0$  &       $2.71\pm0.12$ &       $0.54$ \\
$12.6$  &       $3.82\pm0.21$ &       $0.53$ \\
$13.5$  &       $2.49\pm0.12$ &       $0.54$ \\
$14.2$  &       $1.28\pm0.11$ &       $0.36$ \\
$17.0$  &       $2.52\pm0.27$ &       $1.11$ \\
$17.4$  &       $0.397\pm0.065$ &       $0.21$ \\
$17.9$  &       $0.802\pm0.099$ &       $0.29$ \\
$18.9$  &       $2.19\pm0.14$ &       $0.36$ \\
$33.1$  &       $3.54\pm0.51$ &       $1.66$ 
\enddata
\tablecomments{Only detections $>3\sigma$ are listed. Errors quoted from PAHFIT output.}
\label{table:dustfeatures}
\end{deluxetable}

\begin{deluxetable}{lcccc}
\tablecaption{Additional Infrared Flux Densities from Spitzer}
\tablehead{\colhead{Instrument/Channel} & \colhead{$\lambda$ ($\mu m$)} & \colhead{F$_{\nu}$ (Jy)} & \colhead{$\sigma$ (Jy)} & \colhead{Aperture ($\arcsec$)}}
\startdata
IRAC-2	& $4.5$     & $0.010$ & $0.003$ & $12.2$ \\
IRAC-4	& $8$       & $0.054$ & $0.013$ & $12.2$ \\
MIPS-70	& $70$      & $2.20$ & $0.11$ & $30$ \\
MIPS-160	& $160$     & $0.668$ & $0.033$ & $48$
\enddata
\label{table:spitzer}
\end{deluxetable}

\citet{Spoon07} developed a diagnostic diagram using the $6.2~\mu$m polycyclic aromatic hydrocarbon (PAH) and the strength of the $9.7~\mu$m silicate feature (S$_{sil}$), which is seen in absorption in Cygnus A. These two spectral features can be used in tandem to classify the relative dominance of PAH emission, continuum emission, and silicate absorption. We follow their method of spline fitting to measure the depth of the silicate feature, finding S$_{sil}\approx-0.8$, where S$_{sil}$ is the negative of the apparent optical depth. The EQW of the  $6.2~\mu$m PAH is $0.0552~\mu$m, placing Cygnus A on the border of region 1A and 2A in their diagnostic diagram, corresponding to objects dominated by continuum emission in the mid-infrared. This is consistent with the presence of a strong AGN.

\subsection{Archival Spitzer IRAC+MIPS Data}

The nucleus and hotspots were imaged using IRAC at $4.5$ and $8.0~\mu$m \citep[see ][for a presentation of the data and analysis of hotspot properties]{Stawarz07}. As fluxes for the core were not presented, the archival data were retrieved, re-reduced and calibrated according to the IRAC instrument manual. The images showed strong emission at the location of both radio hotspots and the radio core in both channels. The emission was unresolved in the core component, the measured flux densities for this component are given in Table \ref{table:spitzer}. Extraction apertures of $12\arcsec.2$ were used for both channels. 

\citet{Shi05} presented observation of Cygnus A using the MIPS instrument on Spitzer at 24, 70, and 160 $\mu$m. Their 24 $\mu$m flux is consistent with our IRS observations. The slope between their 70 and 160 $\mu$m points is steeper than that of the Rayleigh--Jeans tail. The 70 and 160 $\mu$m data were re-reduced from the BCD products in the Spitzer Science Center archive. Our measured flux densities are given in table \ref{table:spitzer}. The core of Cygnus A was unresolved in both the $70$ and $160~\mu$m bands. Apertures of $30\arcsec$ and $48\arcsec$ were used at 70 and 160 $\mu$m, respectively. Our 160 $\mu$m measurement has larger error bars, but is broadly consistent with their value.

\subsection{Published Data}

\begin{figure}[h!]
\includegraphics[width=0.5\textwidth]{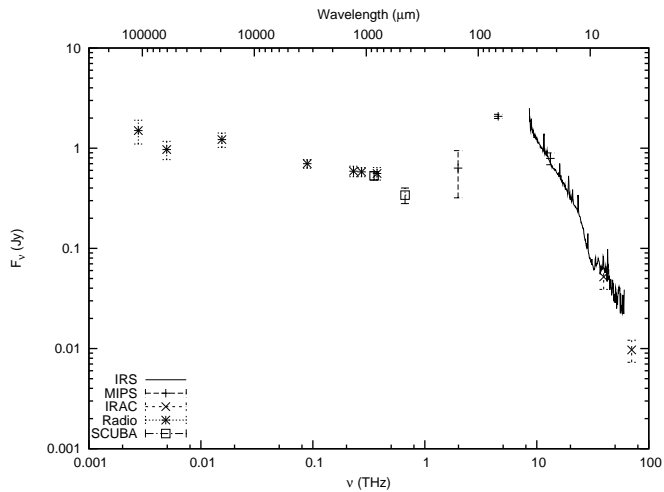}
\caption{Radio through mid-infrared SED with new Spitzer observations and other data from the literature. See the text for details and references.}
\label{fig:SED}
\end{figure}

Cygnus A's strong synchrotron emission at radio frequencies suggests that synchrotron may contribute to the infrared as well. To anchor the synchrotron spectrum we supplemented the Spitzer observations with radio measurements from the literature. Core fluxes were obtained from \citet{Eales89,Salter89,Alexander84,Wright84}. Submillimeter nuclear fluxes were also taken from \citet{Robson98}. Based on these archival data the unresolved radio core\footnote{Here ``radio core'' refers to the core seen by the VLA with on kpc-scale resolution. This encompasses flux from the VLBI scale core and jet.} is flat spectrum ($\alpha=0.18$, $F_{\nu}\propto \nu^{-\alpha}$), until $\sim1$ THz, where thermal emission from dust begins to dominate the SED. The compiled radio through infrared SED is shown in Figure \ref{fig:SED} and the flux densities taken from the literature are provided in Table \ref{table:literature}. The resolution of the observations is also provided.

There is some uncertainty associated with the highest frequency sub-mm observations. The $450~\mu$m observation suggests that the synchrotron break may be occurring in this spectral regime. It is unclear if this is a genuine break or if the measurements are affected by variability. Additional observations at these frequencies may be able to clarify this issue. 

A consideration of the apertures is important when assembling data across several orders of magnitude in frequency. The resolution of the data used to assembled the SED in Figure \ref{fig:SED} varies by over a factor of 10, but in all cases the core component is unresolved. At lower frequencies the emission is due solely to synchrotron emission from the flat spectrum radio core. There is no evidence to suggest that the larger scale steep spectrum jet will contribute emission in the infrared. Accordingly we use the unresolved VLA-scale core to anchor the synchrotron emission in the nucleus, so the difference in apertures should not affect the construction of this SED for the nucleus.

\begin{deluxetable}{lcccc}
\tablecaption{Flux densities from the literature}
\tablehead{\colhead{$\lambda$ ($\mu m$)} & \colhead{F$_{\nu}$ (Jy)} & \colhead{$\sigma$ (Jy)} & \colhead{Resolution ($\arcsec$)} & \colhead{Ref}}
\startdata
450 & 0.34 & 0.06 & $8$ & 1\\
800 & 0.56 & 0.08 & $13$ & 2 \\
850 & 0.53 & 0.05 & $14$ & 1\\
1100 & 0.58 & 0.06 & $19$ & 2 \\
1300 & 0.59 & 0.07 & $11$ & 3 \\
$3.3\times10^3$ (89 GHz) & 0.70 & 0.07 & $2$ & 4\\
$19.5\times10^3$ (15.4 GHz)& 1.22 & 0.20 & Not provided in ref & 5 \\
$60.6\times10^3$ (5 GHz)& 0.97 & 0.20 & $2.0\times3.1$ & 5\\
$109\times10^3$ (2.7 GHz) & 1.5 & 0.4 & $3.7\times5.8$ & 5
\enddata
\label{table:literature}
\tablerefs{1 - \citet{Robson98}, 2 - \citet{Eales89}, 3 - \citet{Salter89}, 4 - \citet{Wright84}, 5 - \citet{Alexander84}.}
\end{deluxetable}

\section{Modeling}
\label{sec:modeling}

We aim to reproduce the major features in the SED: the powerlaw emission at radio wavelengths, the strong thermal emission at infrared wavelengths, and the overall character of the silicate absorption. Extrapolating the powerlaw from the radio to the infrared requires a modification of the powerlaw spectrum to avoid exceeding the observed mid-infrared flux. The infrared emission is thermal in nature, coming from dust at a variety of temperatures ranging from the cold ISM ($T\sim20$ K) through hot dust near the AGN, up to the sublimation temperature ($T\sim1500$ K). The power source for this dust heating is a combination of star formation and AGN activity, with an uncertain balance between the two.

As noted in the introduction previous studies of Cygnus A have found evidence for simultaneous ongoing AGN activity and star formation, both of which can contribute to the infrared emission. We model the continuum emission from $\sim2-10^5$ GHz ($3\times10^{6}-5$ $\mu$m; see Figure \ref{fig:SED}) using components to represent the AGN torus model, a starburst, and synchrotron emission. The selection of models has $15$ free parameters, and the particular choices are discussed in the following sub-sections.

\subsection{AGN Torus Model}

According to the unified scheme for radio loud AGN, the SMBH and accretion disk can be hidden from view along some lines of sight by an obscuring torus. Along these obscured lines of sight the UV/optical radiation is absorbed and re-radiated in the mid- and far-infrared. As noted above, Cygnus A shows evidence for a hidden BLR through observations of H$\alpha$ in polarized light. This suggests that Cygnus A harbors a ``hidden'' accretion disk and a BLR which is obscured along our line of sight.

\citet{Nenkova08} have constructed a model for an obscuring AGN torus, where the obscuration is due to the presence of multiple clouds along the line of sight to the AGN. We select this set of models because clumpy models seem to provide better fits to Sil features than smooth dust distributions \citep[e.g.][]{Baum10}. In order to reproduce the observed ratio of Type I/II AGNs, the obscuring clouds collectively populate a rough toroidal structure with some opening angle. 

\begin{figure}[h!]
\includegraphics[width=0.5\textwidth]{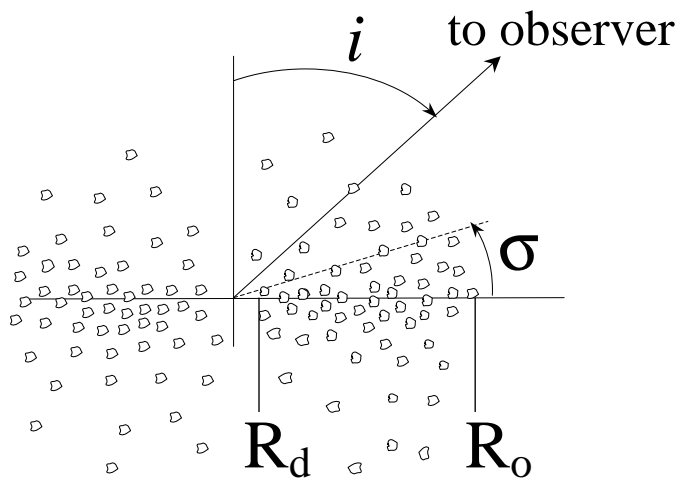}
\caption{Geometry of the CLUMPY torus model. Figure from \citet{Nenkova08}, used with permission. See the text for an explanation of the labels.}
\label{fig:clumpygeom}
\end{figure}

The CLUMPY torus model is specified by multiple geometrical parameters. The outer radius of the torus ($R_o$) is $Y$ times the inner radius ($R_d$), where the inner radius is determined from a dust sublimation temperature of $T=1500$ K ($R_d=0.4\times L_{45}^{0.5}$ pc. $L_{45}$ is the bolometric luminosity of the AGN in units of $10^{45}$ erg s$^{-1}$). The geometry of the model is shown in Figure \ref{fig:clumpygeom}. The clumps have a Gaussian angular distribution,  with $\sigma$ parameterizing the width of the angular distribution from the mid-plane. The radial distribution is a power law with index $q$: $r^{-q}$. The inclination of the torus symmetry axis to the line of sight is $i$ and the average number of clouds along a given line of sight is $N$.

In addition to the geometrical parameters, the models also vary the optical depth of each clump ($\tau_V$), using the \citet{Ossenkopf92} dust composition and \citet{Mathis77} grain size distribution. The overall scaling of the model flux is $F_{AGN}$, the bolometric flux of the AGN's accretion disk component (treating the synchrotron emission separately). 

The clumpy nature of the obscuration implies that there is a finite probability of a direct line of sight to the BLR, even at high inclinations. However, as Cygnus A shows no evidence for directly observed broad lines, we have only fit models where the central regions are fully obscured along our line of sight.

\subsection{Starburst}

Star formation is represented using the \citet{Siebenmorgen07} models which assume spherical symmetry and an ISM with dust properties characteristic of the Milky Way. Emission is broken down into two components: an old stellar population uniformly distributed through the volume and hot, luminous O and B stars embedded in dusty hot spots. The density of OB stars is centrally peaked although the model output is the emission integrated over the entire starburst.

The free parameters for the model are: starburst radius $r$, total luminosity $L_{SB}$, ratio of luminosity in O and B stars to total luminosity $f_{OB}$, visual extinction from the center to the edge of the nucleus $A_V$, and dust density in hotspots around O and B stars $n$. 

Owing to the spherical symmetry of this starburst model and the known ring morphology of the star formation in Cygnus A, parameters such as $A_V$ and the radius $r$ do not have straight forward interpretations here. The assumption of optically thick star formation does not have strong evidence (either for or against), given the absence of high resolution far infrared observations.

\subsection{Synchrotron}
\label{sec:sync-model}

The strong radio emission in Cygnus A justifies the final model component. VLA core fluxes are consistent with powerlaw emission to the submillimeter where thermal emission begins to dominate. Extrapolating the powerlaw to higher frequencies suggests that the synchrotron spectrum must either be modified or subjected to attenuation. The AGN is known to sit behind a dust lane with significant extinction \citep[$A_V=50\pm30$;][]{Djorgovski91}. Though some attenuation of the synchrotron flux is expected at higher frequencies, this alone is not sufficient to explain why the synchrotron powerlaw does not continue through the mid-infrared. With extinction alone, the flux densities at 30 and $10~\mu$ m would be similar for the observed spectral index of $\alpha=0.18$. For $A_V=50$, the expected flux from the extrapolated synchrotron emission would by itself exceed the measured infrared flux at 10$\mu m$. We conclude that there must be a break in the population of emitting electrons.

One possible explanation is a simple cutoff in the population of relativistic electrons at some energy (Case I). This would manifest itself as a cutoff in the synchrotron spectrum:

\begin{equation}
F_{\nu} \propto \nu^{-\alpha_1} e^{-\frac{\nu}{\nu_c}}e^{-\tau_r(\nu)}
\label{eq:synchrotron}
\end{equation}

where $\nu_c$ is the frequency corresponding to the cutoff in the energy distribution of the particles, $\alpha_1$ is the spectral index in the optically thin regime, and $\tau_r(\nu)$ is the dust screen between the synchrotron emitting region and the observer. $\tau_r(\nu)$ follows \citet{Draine84}.

In Cygnus A the spectral index $\alpha_1$ is measured from VLA core radio fluxes, and the amplitude of the powerlaw fixed from the same observations. The only free parameters are the cutoff frequency $\nu_c$ and the optical depth $\tau_{r}$. These parameters art partly degenerate in that in trial runs we experienced a situation where $\tau_r$ and $\nu_c$ would both increase, effectively offsetting each other. To combat this we limited $\tau_r$ such that $A_V$ does not exceed 500 towards the radio source.

An alternate model for the synchrotron emission at higher frequencies is a broken powerlaw behind a dust screen (Case II). Aging of the population of relativistic electrons results in a broken powerlaw whose spectral index increases for frequencies higher than a break frequency \citep{Kardashev62}. The functional form adopted is:

\begin{equation}
F_{\nu}  \propto e^{-\tau_r(\nu)} \times
  \begin{cases} 
    \nu^{-\alpha_1}  & \text{if } \nu < \nu_{break} \\ 
    \nu^{-\alpha_2}  & \text{if } \nu \geq \nu_{break}
  \end{cases}
\label{eq:brokenpw}
\end{equation}

For a simple aging of the electron population, the post-break spectral index in Cygnus A would be $\alpha_2=1.24$. As with Case I, we fit a dust screen in front of the synchrotron emission ($\tau_r$).

\subsection{Stellar Contribution to the Mid-infrared}

Starlight can potentially contribute to the mid-infrared flux, especially in a large aperture. Using the flux of the stellar component from \citet{Jackson98} and their brightness profile, we determined the contribution due to starlight in a $20\arcsec$ aperture. The relative fluxes were consistent with an elliptical galaxy template from \citet{Silva98}. Expected flux densities at 2.2, 5, and 10 $\mu$m were computed using the same template spectrum. After subtracting a nuclear point source, the K-band flux density in starlight is consistent with the flux in a $20\arcsec$ aperture as measured using Two Micron All Sky Survey (2MASS) data products \citep{Skrutskie06}. At 5 and 10 $\mu$m, the starlight contributes roughly 14\% and 2\% respectively, of the flux measured by IRS. Therefore, we do not include any contribution to the mid-infrared flux from an old stellar population in our modeling.

\subsection{Dust in the NLR}

Warm dust in the NLR directly illuminated by the AGN can also contribute to the mid- and far-infrared emission \citep[e.g.,][]{Groves06}. \citet{Ramos09} find that a significant contribution to the mid-infrared flux can come from sources other than the AGN torus, particularly additional hot dust. In Cygnus A \citet{Radomski02} find an extended component to the mid-infrared emission which is consistent with $T\sim150$ K dust. Some of this extended emission is co-spatial with sites of possible star formation. Dust of this temperature is well reproduced with our choice of starburst models (with the implicit assumption that this dust is heated by star formation). 

Higher temperature dust is also present in the inner regions of Cygnus A. However the resolved mid-infrared images of the nuclear regions by \citet{Radomski02} are unable to distinguish between dust in the ``torus'' and dust in the NLR heated by the AGN. Without strong observational motivation for an additional dust component we opt not to include one. As will be shown in the next section we are able to reproduce the observed emission without this additional component.

\subsection{Prior Constraints on Model Parameters}

Prior to fitting the models, the available parameter space was constrained using results from previous studies of Cygnus A.  The opening angle and covering fraction of the CLUMPY torus are determined from the $\sigma$ and $N$ parameters \citep[see Equations (3) and (4) in ][]{Mor09}. We define the half opening angle of the torus as the angle at which the escape probability of a photon drops below $e^{-1}$:

\begin{equation}
\theta_{half}=90-\sigma\sqrt{\ln{N_0}}
\label{eq:halfopening}
\end{equation}

\citet{Tadhunter99} measured the opening angle of the ionization cone in Cygnus A using near infrared HST data, finding $\theta_{half}=(58\pm4)^{\circ}$. We use this value as the torus opening angle and limit the parameter range of $\sigma$ and $N$ according to Equation \ref{eq:halfopening}.

\begin{figure*}
\includegraphics[angle=270,width=0.5\textwidth]{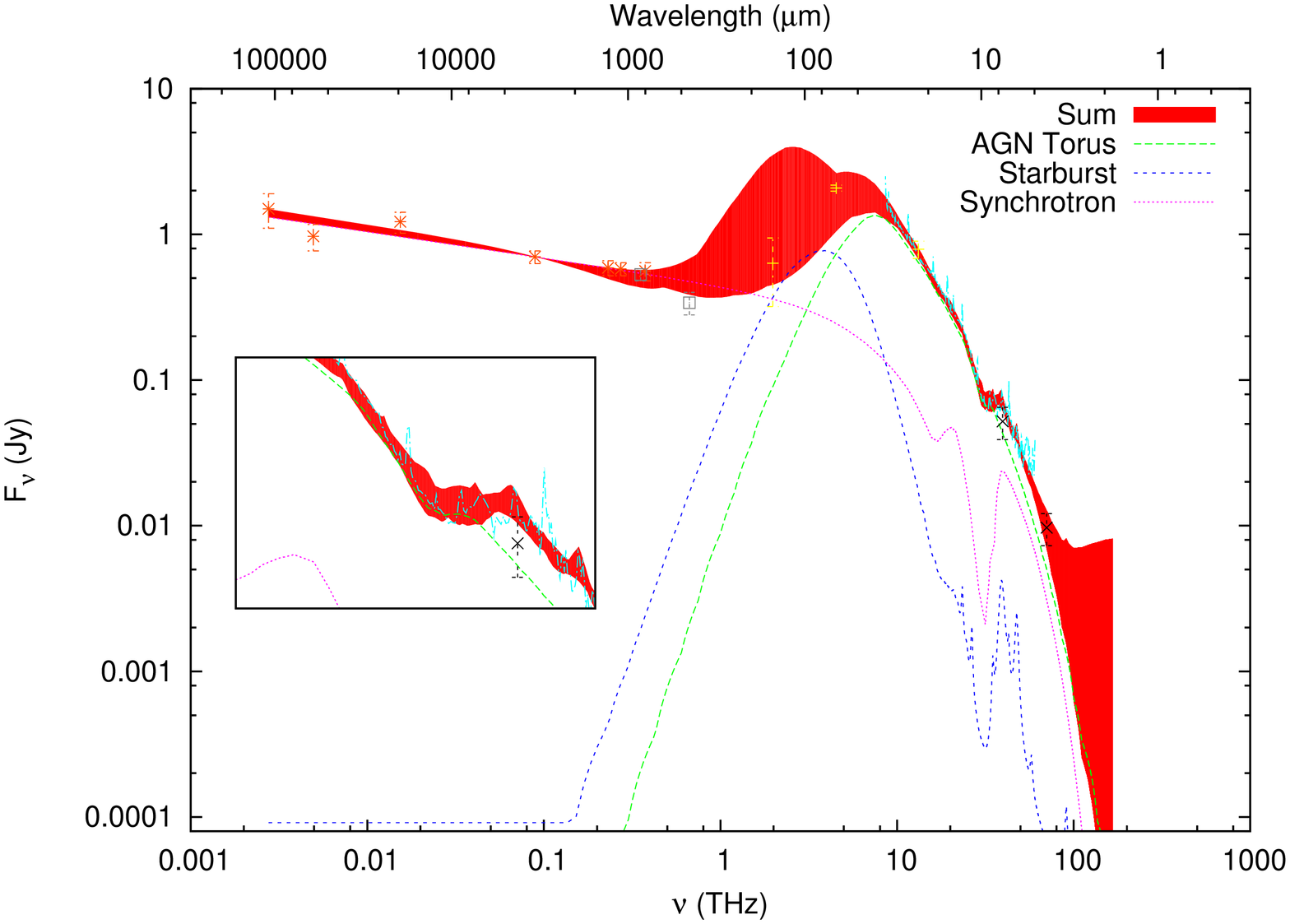}\includegraphics[angle=270,width=0.5\textwidth]{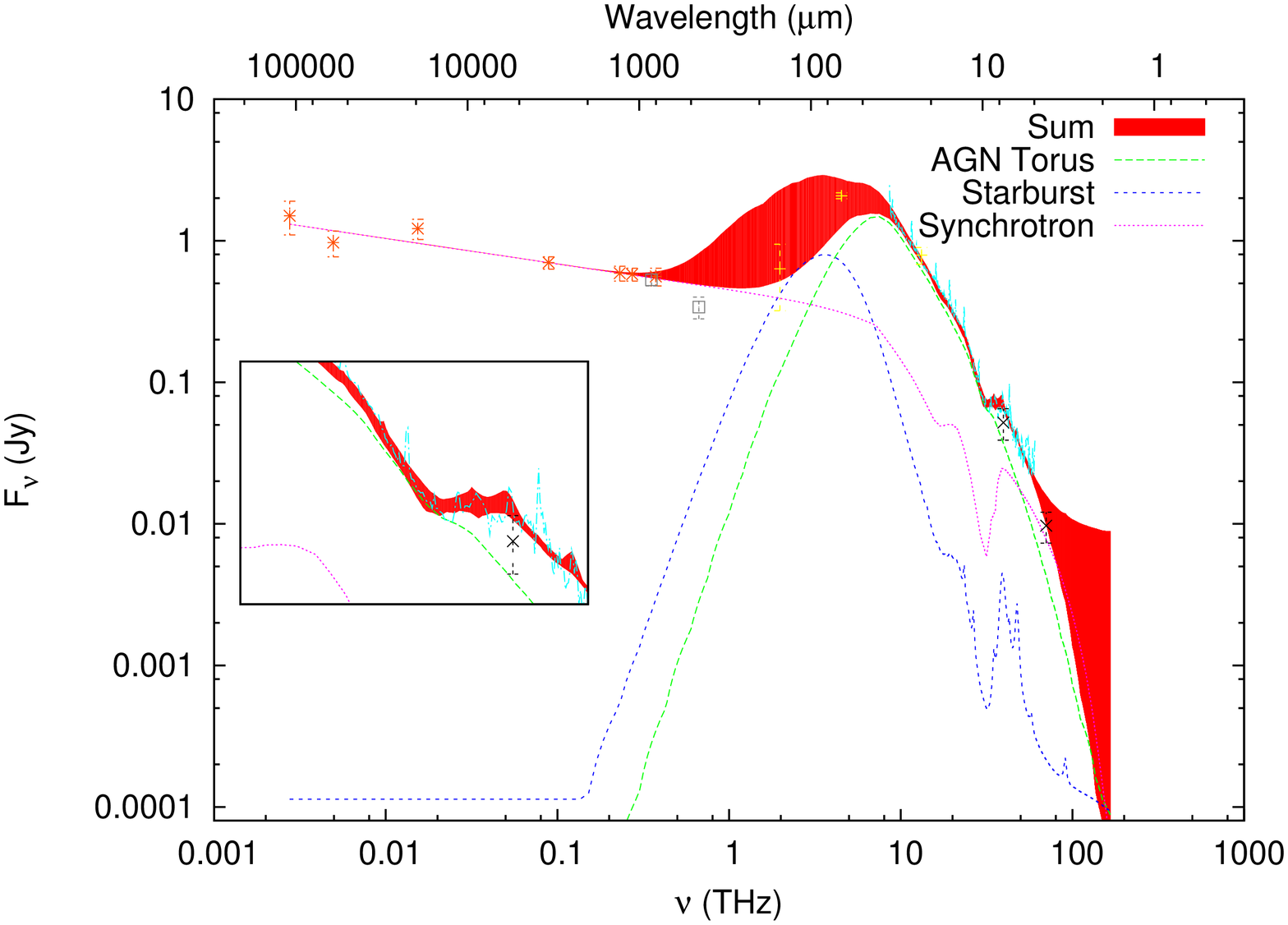}
\caption{LEFT: Case I fit to the Cygnus A SED. The shaded region shows all acceptable model fits within the 95.4\% confidence interval. The lines are the components for the best-fit (lowest $\chi^2$). RIGHT: Same as left, but for Case II fit to the Cygnus A SED. Insets are a zoom of the region around the $9.7~\mu$m Sil absorption.}
\label{fig:SEDfit}
\end{figure*}

\begin{figure*}
\begin{center}
\includegraphics[angle=270,width=0.8\textwidth]{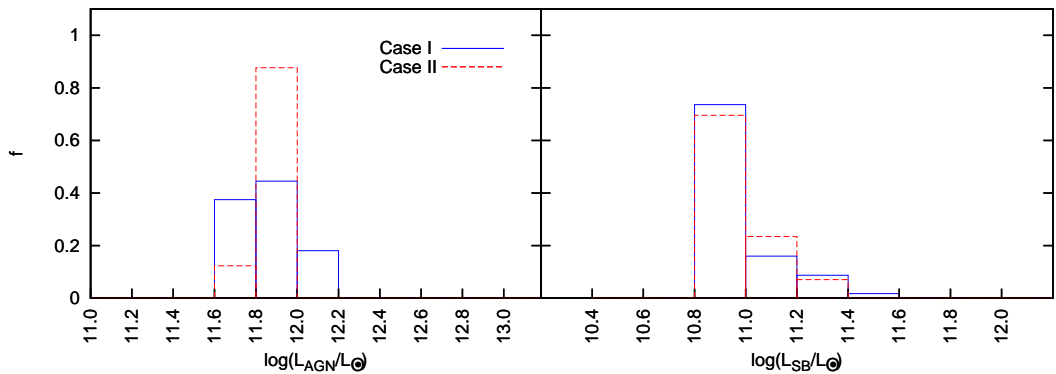}\\
\end{center}
\caption{Histogram of bolometric AGN luminosity from best fit torus parameters (left) and bolometric starburst luminosity (right) for Cases I (blue, solid line) and II (red, dashed line).}
\label{fig:lumhist}
\end{figure*}

The inclination range of the torus $i$ was limited by very long baseline interferometry (VLBI) observations and modeling of the inner pc-scale jet \citep[$50^{\circ}\leq i \leq 85^{\circ}$,][]{Sorathia96}. Additionally, the available VLA core radio fluxes were used to fix the synchrotron spectrum at radio frequencies ($\alpha_1$, and the amplitude).

\section{Results}
\label{sec:results}

To fit the observed SED, the flux from each model was summed in each wavelength bin and compared to the observed flux within a $20\arcsec$ aperture. Optimization was performed using Levernberg--Marquardt least-squares minimization.

Some degeneracy between parameters was seen in the model results. Comparison model runs with fewer prior constraints on the parameters resulted in similar $\chi^2$ values to those quoted below for regions of parameter space which are unlikely to be physically reasonable matches to Cygnus A (e.g., CLUMPY torus fits with covering fractions of unity and $\theta_{half}=0^{\circ}$). Prior constraints from other observations are thus important in eliminating regions of parameter space which may be statistically reasonable but physically unrealistic.

After the models were fit, confidence intervals were determined separately for Case I and Case II through bootstrapping with replacement \citep{Efron81}. These were used to determine the ``acceptable'' range of parameter values.

\begin{figure*}
\includegraphics[angle=270,width=\textwidth]{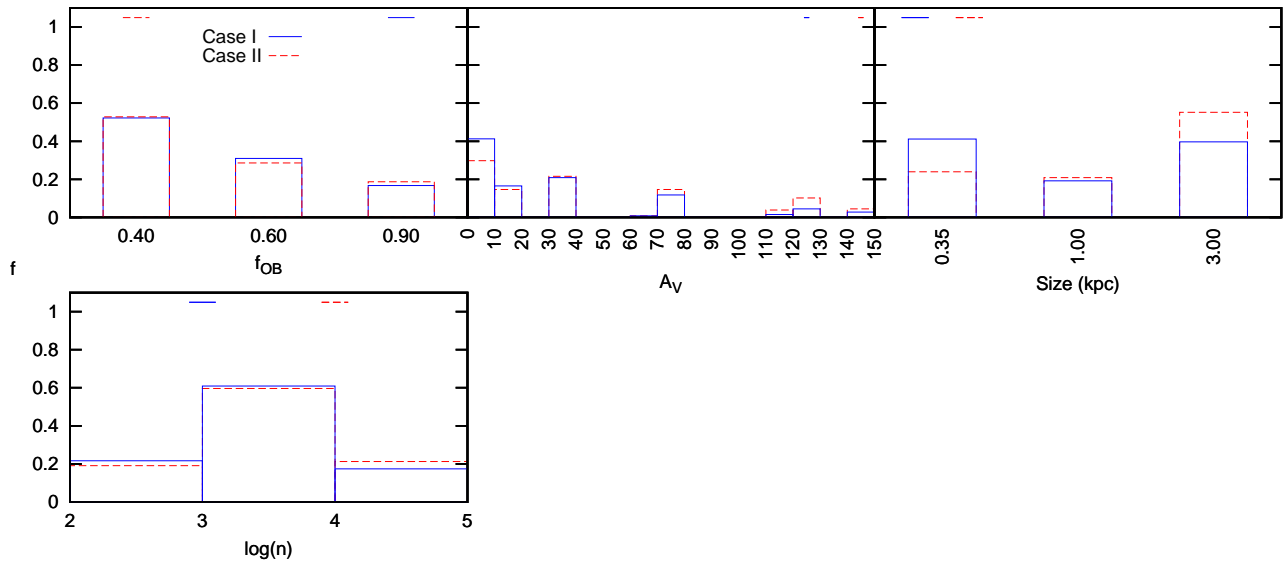}
\caption{Histograms of starburst parameters for acceptable fits within the 95.4\% confidence interval for Cases I (blue, solid line) and II (red, dashed line). The small horizontal bars denote the parameter value for the best-fit model. From left to right, top to bottom: the fraction of luminosity in O and B stars, integrated extinction through the starburst, size of the starburst, dust density in hotspots around O and B stars. f$_{OB}$ and the size are discrete and the histograms represent the relative frequency of models with those specific values. The A$_V$ and $n$ parameters are more finely sampled than the histogram represents; the plots show the relative fraction of models with parameters in the noted ranges.}
\label{fig:sbhist}
\end{figure*}

The results of the fits are shown in Figure \ref{fig:SEDfit}. The range of parameter values for fits within at $95.4\%$ confidence interval are shown in Figures \ref{fig:lumhist}-\ref{fig:synchist}. Section \ref{ssec:sync-exp} discusses the Case I fits (exponential cutoff in the relativistic electron population) and Section \ref{ssec:sync-break} discusses the Case II fits (broken powerlaw for the synchrotron emission).

\subsection{Case I: Synchrotron Exponential Cutoff}
\label{ssec:sync-exp}

The best fit combination of parameters for Case I has $\chi^2$/DOF = $1.35$. The range of acceptable matches for the 1641 model combinations is shown in Figure \ref{fig:SEDfit} (left). Histograms of the best fit parameters are given in Figures \ref{fig:lumhist} (left) and \ref{fig:agnhist}, where $f$ is the fraction of models within the 95.4\% confidence interval in a given bin. Parameters for the best fit model are marked with a small blue horizontal bar.

\subsubsection{AGN/Torus Properties and Contribution}

\begin{figure*}
\includegraphics[angle=270,width=\textwidth]{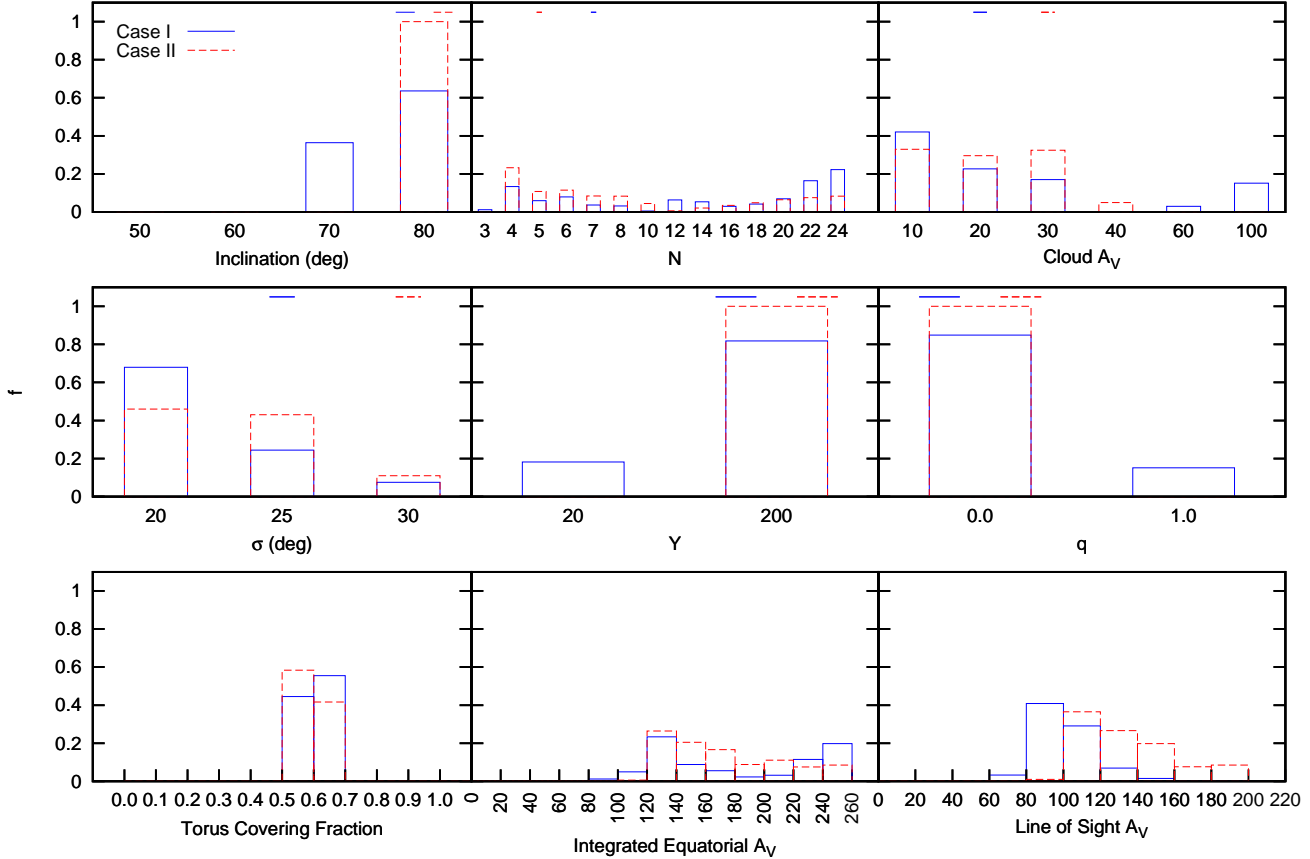}
\caption{Histograms of torus parameters for acceptable fits within the 95.4\% confidence interval for Cases I (blue, solid line) and II (red, dashed line). The small horizontal bars denote the parameter value for the best-fit model. From left to right, top to bottom: viewing angle of the torus ($90$ meaning the torus is viewed edge-on), N - average number of clouds along an equatorial line of sight, A$_V$ - for an individual cloud, $\sigma$ - angular width of the distribution of torus clouds, Y - ratio of inner to outer radii, q - power law index of the radial distribution of torus clouds, covering fraction of the torus (equivalent to the escape probability of optical/UV photons, assuming optically thick clouds), integrated A$_V$ along an equatorial line of sight, integrated A$_V$ along our line of sight (calculated using N, $\sigma$, the cloud A$_V$, and the inclination). Parameters in the first two rows are discrete and the histograms represent the relative number of models with those specific values. Parameters in the bottom row are derived from the parameters in the top two rows as well as the luminosity of the AGN component (shown in Figure \ref{fig:lumhist}), and are continuous.}
\label{fig:agnhist}
\end{figure*}

The fits favor a bolometric accretion disk luminosity of $log(L_{AGN}/L_{\odot})\sim 11.8-12.0$ (median of $11.82$ with an interquartile spread of $0.09$). The fits clearly favor an extended torus ($Y=200$) with low opacity clouds ($A_V\sim10-30$). For the median bolometric luminosity, the inner radius of the torus is $R_d=0.6$ pc, giving a corresponding outer torus edge of $\sim125$ pc (for Y$=200$). 

The range of values for $i$ was limited based on previous work in the radio regime. The fits prefer an inclination for the torus on the high end of the range, $i\approx80^{\circ}$. $N$ has a bimodal distribution, with $25$\% of fits having $N\leq6$ clouds along an equatorial line of sight and $\sim60$\% having $N\geq20$ clouds. Most of the fits within the 95.4\% confidence interval have a flat radial distribution of clouds ($q=0$). 

The opacity of individual clouds is anti-correlated with the radial extent of the torus; small torus sizes favor high $A_V$ values. High values of $N$ (i.e., many clouds along a given line of sight) are favored for small torus sizes, however for large torus sizes, both small and large $N$ values are acceptable.

We calculate the total extinction through the torus along both an equatorial line of sight (Figure \ref{fig:agnhist} middle, bottom row), and along our line of sight to the torus (Figure \ref{fig:agnhist} right, bottom row). The integrated equatorial $A_V$ spans a range between $100$ and $250$. When viewing angle is taken into consideration, the range narrows, with most fits showing the line of sight $A_V$ between $80$ and $120$. However the $A_V$ is poorly constrained beyond having a well-defined lower bound.

The torus covering fraction was computed using the method of \citet[their Equations (3) and (4)]{Mor09}. By design, the derived covering fractions for the torus are between $50\%$ and $70\%$. 

\subsubsection{Starburst}

The median value of the starburst luminosity for Case I is log$(L_{SB}/L_{\odot})\sim10.8$, with a tail up to $\sim11.6$ containing approximately $40\%$ of the fits (Figure \ref{fig:lumhist}). This covers a range of star formation rates from $10$ to $70$ M$_{\odot}$ yr$^{-1}$, as determined by the $L_{IR}$ calibration from \citet{Kennicutt98a} \citep[for a review of SFR estimates, see][]{Kennicutt98b}.

\begin{equation}
\frac{SFR}{(M_{\odot}~yr^{-1})} = 4.5\times10^{-44}~\frac{L_{FIR}}{erg~s^{-1}}=1.72\times10^{-10}~\frac{L_{FIR}}{L_{\odot}}
\label{eq:sfr}
\end{equation}

Roughly half the fits within this confidence interval have $40$\% of the starburst luminosity in the form of OB stars. The distribution of size parameters is relatively flat, although even the largest sizes would be unresolved by our Spitzer observations. The distribution of dust density $n$ peaks around $10^3$ cm$^{-3}$. The extinction through this starburst is relatively unconstrained.

\begin{figure*}
\includegraphics[angle=270,width=\textwidth]{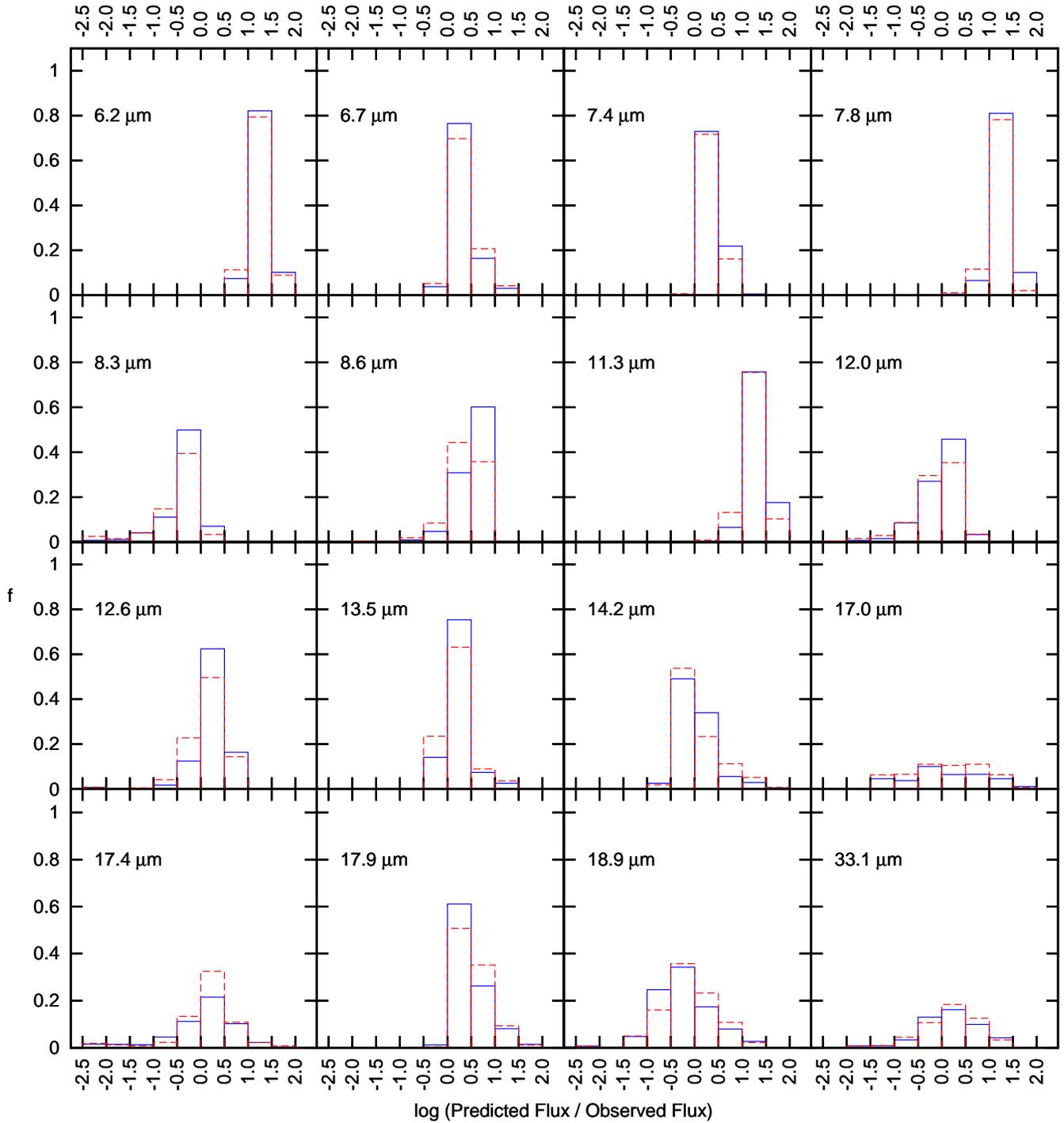}
\caption{Histograms of log(Predicted Flux / Measured Flux) for dust features detected in the IRS spectrum of Cygnus A. The predicted values were measured from the \citet{Siebenmorgen07} models using PAHFIT. The solid blue line denotes Case I while the red dashed line denotes Case II. (see Table \ref{table:dustfeatures} for the measured flux values).}
\label{fig:dust}
\end{figure*}

The strengths of dust features in starburst models falling within the 95.4\% confidence interval were also measured using PAHFIT to compare with the intensities of observed dust feature. Figure \ref{fig:dust} shows histograms of the predicted intensities from our models divided by the measured intensity from the IRS spectrum. Only dust features which are detected in the IRS spectrum have been plotted. In general the agreement is good with most models matching the measured line intensities. However for some dust features a significant number of the models within the confidence interval predict emission in excess of what is observed. While a detailed comparison of the relative strengths of dust features is beyond the scope of the paper, we suggest the models are able to generally reproduce the dust features seen in the spectrum. The dust features which show the greatest discrepancy between observed fluxes and those predicted from the modeling ($6.2$, $7.8$, and $11.3~\mu$m) comprise 3 of the 5 lowest signal-to-noise dust feature fits in the spectrum. The discrepancy may then be due to the difficulty of fitting low equivalent width dust features.

PAHFIT also attempted to fit eight other dust features in the IRS spectrum. The upper limits for three ($5.7$, $14.0$, and $15.9~\mu$m) are consistent with expectations from the starburst models. One dust feature ($16.4~\mu$m) has a measured limit below that expected from the best fitting starburst model, and so is discrepant. The other four features ($7.6$, $10,7$, $11.2$, and $12.7~\mu$m) also have upper limits from PAHFIT which are lower than the expected value from the starburst models. However, these are coincident with or near other spectral features (e.g., narrow emission lines or the Sil absorption). Thus an accurate measurement of these dust features in the IRS spectrum would rely more heavily on fitting the wings, which could prove difficult in a continuum dominated source such as Cygnus A. 

\subsubsection{Synchrotron Radiation}

The synchrotron emission amplitude was fixed using the non-thermal emission from the radio core, assuming it to be a point source at all frequencies observed. The synchrotron model, therefore, has only two parameters: cutoff frequency ($\nu_c$) and extinction due to dust ($\tau_r$).

\begin{figure*}
\includegraphics[angle=270,width=\textwidth]{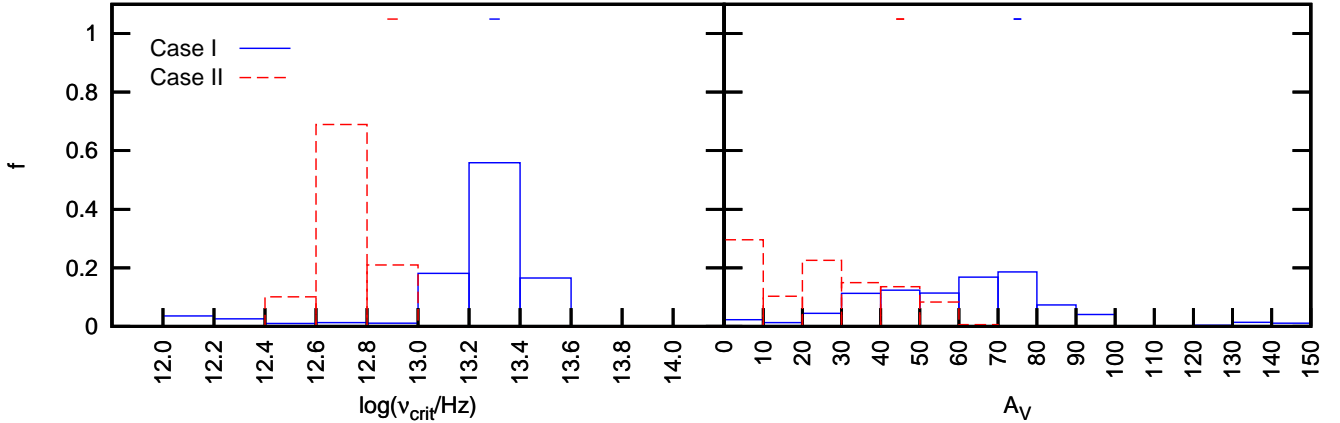}
\caption{Histograms of synchrotron parameters for acceptable fits within the 95.4\% confidence interval for Cases I and II. For Case I $\nu_{crit}\equiv\nu_{cutoff}$ and for Case II $\nu_{crit}\equiv\nu_{break}$. $A_V=1.08\tau_r (\nu= \text{V-Band} )$.}
\label{fig:synchist}
\end{figure*}

The distribution of cutoff frequencies is somewhat broad, covering the range of $\nu\approx10-60$ THz ($5-30~\mu$m) (Figure \ref{fig:synchist}). The fits show a range of acceptable extinction for the dust screen, peaking in the range of $A_V\approx60-80$, slightly lower than the predicted line of sight $A_V$ from the torus models. The integrated luminosity of the ``core'' synchrotron flux is log$(L_{sync}/L_{\odot})\sim11.2$. 

This break in the synchrotron spectrum is consistent other powerful FR II radio sources, where an extrapolation of the radio synchrotron emission significantly exceeds the observed optical flux \citep[e.g.,][]{Schwartz00,Sambruna04,Mehta09}. In these cases, the similarity of the radio and X-ray spectral indices suggests that the X-ray may be produced by inverse Compton (IC) scattering.

With the modeled synchrotron spectrum, a magnetic field strength, and the assumption that each electron emits at a single frequency, the energy distribution of relativistic electrons can be determined:

\begin{equation}
\gamma(\nu)= \sqrt{\frac{4\pi m_e c \nu}{3eB}}
\label{eq:e-gamma}
\end{equation}

where all constants and values are in the cgs system of units.

\clearpage

The magnetic field strength has been estimated from VLBI observations to be roughly $17$ mG ($100$ mG) in the jet (core) \citep{Roland88,Kellermann81}. Using $\nu\approx30$ THz and Equation \ref{eq:e-gamma} we find $\gamma\sim2\times10^4$ ($8\times10^3$) for the jet (core).

In addition to radiating energy via synchrotron emission, relativistic electrons can also lose energy via IC scattering. The ratio of the energy lost through synchrotron to the energy lost to IC is simply given by the ratio of the magnetic and radiation energy densities, with the cosmic microwave background (CMB) setting a minimum value for the radiation energy density. Using the magnetic field strength assumed above ($17$ mG), and $T_{CMB}=2.73$ K, synchrotron losses are dominant by a factor $>10^8$. A larger magnetic field would further enhance this ratio, so IC losses from scattering of CMB photons have a negligible effect on the population of relativistic particles.  

\subsection{Case II: Synchrotron Broken Power Law}
\label{ssec:sync-break}

The range of best fit models for Case II is shown in Figure \ref{fig:SEDfit} (right). The best fit model had $\chi^2$/DOF = $1.03$. Figure \ref{fig:SEDfit} (right) shows the range of the 1580 model fits within the 95.4\% confidence interval. Histograms of the best fit parameters are given in Figures \ref{fig:lumhist} (left) and \ref{fig:agnhist}. A red horizontal bar marks the parameter for the parameters with the lowest $\chi^2$ value.

\subsubsection{AGN/Torus}

In Case II fits the bolometric accretion disk luminosity is roughly log$(L_{AGN}/L_{\odot})\sim 11.8$ (median $11.88$ with an interquartile range of $0.08$). The distribution of luminosities is more strongly peaked than for Case I, with over 80\% of fits occupying the log$(L/L_{\odot})=11.8-12.0$ bin. A large torus ($Y=200$) is exclusively preferred, corresponding to $R_{out}\approx135$ pc (using the median value for the luminosity). Again, by design, the torus covering fraction is between 0.5 and 0.7. An inclination of $i=80^{\circ}$ is exclusively preferred.

The distribution of the average number of clumps along a line of sight $N$ is again bimodal with either large ($<14$) or small ($<8$) numbers of clouds preferred. The number of clouds is anti-correlated with the extinction through individual clouds. The spread in equatorial $A_V$ is somewhat large ($120-260$), but the line of sight $A_V$ is more constrained ($100-160$ for 80\% of fits). 

\subsubsection{Starburst}

The starburst component in Case II fits show qualitatively similar behavior to Case I. The typical luminosity is comparable, but shows a smaller tail up to log$(L_{SB}/L_{\odot})=11.6$ ($\sim30\%$ of fits). Other parameters are similar to those in Case I fits, with the exception of the size which appears to have a very slight preference for a larger size, suggesting an overall cooler dust temperature for a given luminosity. This may indicate a degeneracy between the starburst and synchrotron models. The critical frequency for the synchrotron spectrum (see the next section) is at lower frequency for Case II, resulting in a smaller contribution to the far-infrared flux. The starburst model compensates with a larger overall size to provide additional cool dust emission (for a given starburst luminosity).  The dust feature strengths expressed in the models generally compare favorably with Case I.

\subsubsection{Synchrotron Properties and Contribution}

An alternate mechanism for limiting the influence of synchrotron emission at shorter wavelengths is for the spectrum to break at some frequency (see Section \ref{sec:sync-model}). Case II fits used a fixed pre-break spectral index of $\alpha_1=0.18$ and a post-break spectral index of $\alpha_2=1.24$, consistent with aging of the relativistic population \citep[without injection of additional particles;][]{Kardashev62}. As the flux density decreases in a slower fashion when compared to an exponential cutoff, the powerlaw must break at lower frequencies to ensure that the observed mid-infrared flux is not exceeded. 

The break frequency is roughly 5 THz ($60~\mu$m). The unobscured synchrotron luminosity is found to be log$(L_{sync}/L_{\odot})\sim11.1$. Following the same arguments as Case I, the electrons emitting at the break frequency have $\gamma\sim7\times10^3$ ($3\times10^{3}$) for jet (core) magnetic field values. Again, this is consistent with the observed properties of jets in other FR II radio sources.

The extinction of the dust screen in front of the radio source is lower than Case I fits, with $A_V<60$, still within range of the extinction to the central source estimated by \citet{Djorgovski91}, but lower than the computed line of sight $A_V$ from the torus. The larger discrepancy compared with Case I is due to the enhanced short wavelength emission of the broken power law at shorter wavelengths (when compared to the Case I exponential cutoff). The equatorial torus $A_V$ increases to provide an overall cooler temperature, and the modeled dust screen in front of the synchrotron component remains small (in order to contribute sufficient flux at shorter wavelengths).

\section{Discussion of General SED Results}
\label{sec:discussion}

\subsection{Luminosity and Kinetic Power in Cygnus A}

Our best estimate of the bolometric AGN luminosity in Cygnus A (log$(L/L_{\odot})\sim12$, including the synchrotron component and X-ray emission) is above the \citet{Whysong04} estimate using Keck mid-infrared observations and a PG-quasar spectrum ($3.9\times10^{11}$ $L_{\odot}$), but is somewhat below the estimates of \citet{Tadhunter03} who find $L_{bol}=12-55\times10^{11}$ $L_{\odot}$ (although consistent with the low end of their range). The bolometric AGN luminosity is the best constrained parameter, and is insensitive to the synchrotron model adopted.

The kinetic power in the expansion of the radio lobes in Cygnus A can be inferred from X-ray observations of the cluster environment \citep{Wilson06}. From this analysis of Chandra observations Cygnus A has a kinetic power of $L_{kin} = 1.2 \times 10^{46}$ erg s$^{-1}$ (log$(L_{kin}/L_{\odot})=12.5$), a factor of three larger than the bolometric AGN luminosity inferred from the modeling. Given the uncertainties in the determinations of both the kinetic and bolometric power, it is unclear if the kinetic power dominates significantly over the power emitted as radiation.

\subsection{Implications of the Modeling}

The probable torus sizes from the SED modeling give outer radii of $R_o=130$ pc ($\approx0\arcsec.2$ at the distance of Cygnus A). Generally, the torus parameters provide reasonable estimates of the properties of the obscuring structure. This predicted outer radius is significantly larger than what is generally assumed to be reasonable for the obscuring torus. A torus with the angular diameter of the $\approx0.\arcsec2$ is well within range of multiple mid and near-infrared instruments. At large radii, the distinction between torus clouds and narrow line region clouds may be somewhat arbitrary. The observations by \citet{Canalizo03} show extended emission on scales including and larger than our fit torus sizes. The degree of contamination by emission lines is unclear, and it would be interesting to attempt to obtain a line-free continuum image to ascertain the true size and shape of the continuum emitting region.

The torus size parameters are consistent with those found from modeling of the $9.7~\mu$m Sil feature by \citet{Imanishi00}. Their radiative transfer modeling suggested some torus properties similar to those presented here, namely a small inner radius ($<10$ pc) and an inner-to-outer radius of $80-500$. In contrast they find a steeper radial dependence for the dust distribution $q\sim2-2.5$, possibly due to the use of a smooth dust distribution.

The preferred parameters for the obscuring torus; low $q$, high $A_V$, and large $Y$ suggest the model is being driven toward a cooler spectrum. This could influence the starburst component, leading to a lower inferred star formation rate. In order to combat this it would be beneficial to place tighter constraints on the far-infrared SED to enable better modeling of the star formation in Cygnus A.

A comparison of our IRS data with mid-infrared observations by \citet{Radomski02} is consistent with the suggestion of our model that at $10$ and $18~\mu$m, the emission is dominated by the torus+synchrotron component. \citet{Tadhunter99} measure a K-band nuclear point source with $F_{\nu}=(49\pm10)~\mu$Jy. This flux limit is broadly consistent with some model fits for the torus+synchrotron component.

For the estimated bolometric luminosity $10^{12}$ $L_{\odot}$, and a black hole mass of $2.5\times10^9$ $M_{\odot}$, the Eddington Ratio ($L/L_{edd}$) is $\sim1.3\times10^{-2}$. This is similar to, but slightly lower than previous estimates from \citet{Tadhunter03}, likely due to the fact that our model attributes some of the IR luminosity to the heating of dust from star formation. This starburst luminosity is relatively well constrained, with consistent values across the bulk of fits. The AGN (torus plus jet) contributes $\sim90\%$ of the infrared luminosity and star formation produces the remaining $10\%$.

\subsection{Future Work and Observations}

The results of our modeling can be tested and improved with the help of future observations in various wavelength regimes.

Around 1 THz ($\sim300\mu$m) the synchrotron and starburst model components are of similar flux density, with the synchrotron contribution decreasing and the thermal contribution from star formation beginning to dominate. Unfortunately, data at this location are limited in resolution, sensitivity, and wavelength coverage. Disentangling the emission from the starburst and emission from the AGN at this frequency will be possible with the Atacama Large Millimeter Array \citep[ALMA; e.g.,][]{Carilli05}, particularly with the availability of Band 10 and full science capabilities.

The far-infrared suffers from poor sampling of the SED. In this regime the Herschel Space Observatory \citep{Pilbratt10} provides an opportunity to improve on the understanding of the continuum emission. SPIRE and PACS cover wavelength ranges of interest: the region of the possible jet-break and the long wavelength side of the infrared bump. Observations here can provide improved constraints for the synchrotron and starburst models. \citet{Gonzales10} demonstrate decomposition of far-infrared emission for Mrk 231, with the Herschel SPIRE observations providing important constraints. Similar observations of Cygnus A will provide important data on this sparsely observed portion of the SED. PACS observations would contribute measurements of far-infrared fine structure lines which could be used in concert with the mid-infrared lines to provide an alternate method of determining the relative contribution of star formation and AGN activity to the infrared luminosity \citep[e.g.,][for Mrk 231]{Fischer10}.

Although the synchrotron spectrum breaks in our fits, it may still dominate the flux between $5$ and $10~\mu$m. This result can be tested by future mid-infrared polarimetry or variability studies. Synchrotron emission from a compact source such as the radio core or jet knots should be subject to flaring. If the emission in this wavelength range is instead of a thermal origin (e.g., hot dust in the host galaxy), the $10~\mu$m flux will remain relatively stable.

\section{Summary}
\label{sec:summary}

Using a combination of a new mid-infrared spectrum from the Spitzer Space Telescope and radio data from the literature, the nuclear emission in Cygnus A has been modeled as a combination of powerlaw emission from a synchrotron jet, reprocessed AGN emission from a dusty torus, and emission from a dusty circumnuclear starburst.

The data are well fit by a combination of these three models, and all three are necessary to reproduce the observed emission. Statistically acceptable fits were found for both an exponential cutoff in the population of relativistic electrons (Case I) as well as emission from an aging electron population (Case II), however we are unable to distinguish between the two cases. For Case I we find the cutoff frequency to be between $10$ and $50$ THz ($5-30~\mu$m) while for Case II fits the predicted break frequency is $5$ THz ($60~\mu$m). Degeneracy between the starburst and synchrotron components makes a more precise determination of the break or cutoff frequency difficult. Better observations on the long-wavelength side of the thermal bump will provide tighter constraints on the models.

From this modeling, we find the following:
\begin{itemize}
  \item{The bolometric luminosity of the AGN in Cygnus A is $\sim10^{12}$ $L_{\odot}$}
  \item{The mid-infrared emission is consistent with emission from a clumpy obscuring torus with an outer size of $\sim130$ pc.}
  \item{The far-infrared emission is consistent with being dominated by star formation which is occurring at a rate between $10$ and $70$ $M_{\odot}$ yr$^{-1}$.}
\end{itemize}

In Cygnus A, the infrared emission is a combination of AGN and starburst heated dust, with the AGN contributing $\sim90\%$ of the luminosity.

\acknowledgements

The authors thank the anonymous referee who's comments have improved the quality of this paper. 

This work is based on observations made with the Spitzer Space Telescope, which is operated by the Jet Propulsion Laboratory, California Institute of Technology under a contract with NASA. Support for this work was provided by NASA through an award issued by JPL/Caltech. Computing resources were provided in part by the Research Computing group of the Rochester Institute of Technology. This research has made use of the NASA/IPAC Extragalactic Database (NED) which is operated by the Jet Propulsion Laboratory, California Institute of Technology, under contract with the National Aeronautics and Space Administration. This research has made use of NASA's Astrophysics Data System. This work is supported in part by the Radcliffe Institute for Advanced Study at Harvard University.

G.P. thanks D. Whelan, D. M. Whittle, and A. Evans for helpful discussions.

\end{document}